\documentstyle[prl,floats,aps,twocolumn,epsf,graphicx]{revtex}
\begin{document}
\twocolumn[\hsize\textwidth\columnwidth\hsize\csname
@twocolumnfalse\endcsname

\title{Selection Rules for Black-Hole Quantum Transitions}
\author{Shahar Hod$^1$ and Uri Keshet$^2$}
\address{$^1$The Racah Institute of Physics, The Hebrew University, Jerusalem 91904, Israel}
\address{}
\address{$^2$Physics Faculty, Weizmann Institute, Rehovot 76100, Israel}
\date{\today}
\maketitle

\begin{abstract}

\ \ \ We suggest that quantum transitions of black holes comply with selection rules, analogous to those of atomic spectroscopy. In order to identify such rules, we apply Bohr's correspondence principle to the quasinormal ringing frequencies of black holes. In this context, classical ringing frequencies with an asymptotically {\it vanishing} real part $\omega_R$ correspond to virtual quanta, and may thus be interpreted as forbidden quantum transitions. With this motivation, we calculate the quasinormal spectrum of neutrino fields in spherically symmetric black-hole spacetimes. It is shown that $\omega_R\rightarrow 0$ for these resonances, suggesting that the corresponding fermionic transitions are quantum mechanically forbidden. 
\end{abstract}
\bigskip

]

The necessity of a quantum theory of gravity was already 
recognized in the early days of quantum mechanics and general relativity. 
However, despite the flurry of research in this field we still lack a complete theory of quantum gravity.
In many respects the black hole plays the same role in gravitation that the atom 
played in the nascent of quantum mechanics \cite{Bekenmar}. 
It is therefore believed that black holes may play a major role in our 
attempts to shed light on the nature of a quantum theory of gravity.

The quantization of black holes was proposed long ago by Bekenstein
\cite{Beken1,Beken2}, based on the remarkable observation that the horizon
area of a non-extremal black hole behaves as a classical 
adiabatic invariant. In the spirit of the Ehrenfest principle
\cite{Ehren} -- any classical adiabatic invariant
corresponds to a quantum entity with a discrete spectrum, and based on the idea of a minimal increase in black-hole surface area \cite{Beken1}, Bekenstein conjectured that the horizon area of a quantum
black hole should have a discrete spectrum of the form

\begin{equation}\label{Eq1}
A_n=\gamma {\ell^2_P} \cdot n\ \ \ ;\ \ \ n=1,2,3,\ldots\ \  ,
\end{equation}
where $\gamma$ is a dimensionless constant, and 
$\ell_P=(G\hbar/c^3)^{1/2}$ is the
Planck length (we use gravitational units in which $G=c=1$ henceforth). 
This type of area quantization has since been reproduced based on various other considerations (see e.g., \cite{Hod1} for a detailed list of references). 

Mukhanov and Bekenstein \cite{Muk,BekMuk,Beken3} have suggested an independent 
argument in order to determine the value of the coefficient $\gamma$. 
In the spirit of the Boltzmann-Einstein formula in
statistical physics, they relate $g_n \equiv \exp[S_{BH}(n)]$ to the number of the black hole microstates that correspond to a particular external macro-state, where $S_{BH}$ is the black-hole entropy. 
In other words, $g_n$ is the degeneracy of the $n$th area eigenvalue. 
Now, the thermodynamic relation between black-hole surface area and entropy, $S_{BH}=A/4\hbar$, 
can be met with the requirement that $g_n$ has to be an integer for every $n$ only when
 
\begin{equation}\label{Eq2}
\gamma =4\ln{k} \  ,
\end{equation}
where $k$ is some natural number. 

Identifying the specific value of $k$ requires further input. This information 
may emerge by applying {\it Bohr's correspondence principle} to 
the (discrete) quasinormal mode (QNM) spectrum of black holes \cite{Hod2}. 
Gravitational waves emitted by a perturbed black hole are dominated by this 
`quasinormal ringing', damped oscillations with a {\it discrete}
spectrum (see e.g., \cite{Nollert1} for a detailed review). 
At late times, all perturbations are
radiated away in a manner reminiscent of the last pure dying tones of
a ringing bell \cite{Press,Cruz,Vish,Davis}. 
These black-hole resonances are the characteristic `sound' of
the black hole itself, depending on its parameters: mass, charge and angular momentum.

It turns out that for a Schwarzschild black hole, for a given angular harmonic index $l$ there exist 
an infinite number of (complex) quasinormal frequencies, characterizing oscillations with decreasing relaxation times (increasing imaginary part) \cite{Leaver,Bach}. 
On the other hand, it was found numerically  \cite{Leaver,Nollert2,Andersson} that 
the real part of the Schwarzschild gravitational resonances approaches an 
asymptotic constant value (we normalize $2M=1$, and assume a time dependence of the form $e^{-i\omega t}$),

\begin{equation}\label{Eq3}
\omega_n=0.0874247-{i \over 2} \left(n+{1 \over 2} \right)\  .
\end{equation}
Based on Bohr's correspondence principle, it was suggested \cite{Hod2} that 
this asymptotic real value actually equals $\ln(3)/(4\pi)$ \cite{Note1}. 
An analytical proof of this equality was later given in \cite{Motl}. 
This was followed by a calculation of the asymptotic QNM frequencies of scalar 
and gravitational-electromagnetic fields in the charged Reissner-Nordstr\"om (RN) spacetime \cite{MotNei}.

The emission of a quantum of frequency $\omega$ results in a change $\Delta M=\hbar \omega_R$ in the black-hole mass. Assuming that $\omega$ corresponds to the asymptotically damped limit Eq.~(\ref{Eq3}) \cite{Note1}, this implies a change $\Delta A=32\pi M\Delta M=4\hbar \ln3$ in the black hole surface area. 
Thus, the correspondence principle, as applied to the black-hole resonances, 
provides the missing link, and gives evidence in favor of the value $k=3$. 
The coefficient $\gamma =4 \ln 3$ is a unique value, consistent with the area-entropy 
thermodynamic relation, with statistical physics arguments (namely, 
the Boltzmann-Einstein formula), and with Bohr's correspondence principle \cite{Note2}.

Furthermore, it was later suggested to use the black-hole 
QNM frequencies in order to fix the value of the Immirzi parameter in Loop Quantum Gravity, 
a viable approach to the quantization of general relativity \cite{Dreyer,Ash,Rov}. 
The intriguing proposals outlined above \cite{Hod2,Dreyer} have triggered a flurry of research 
attempting to calculate the asymptotic ringing frequencies of various types of black holes (for a detailed list of references see, e.g., \cite{HodKesh}).

The discrete black-hole mass (area) spectrum implies
a {\it discrete} line emission from a quantum black hole; the frequencies of the radiation quanta emitted by the black hole will be integer multiples of the fundamental frequency $\omega_0=\ln3/4\pi$ \cite{BekMuk}. 
If true, this result indicates a distortion of Hawking's semiclassical spectrum \cite{Bekenmar}. 
Such a modification should not be met with surprise: one should bear in mind that 
Hawking's prediction of black-hole evaporation is semiclassical in the sense that the matter fields are treated quantum mechanically, but the spacetime (and the black hole itself)
are treated classically. One therefore suspects that some modifications in the character of the radiation will arise when quantum properties of the {\it black hole} itself are properly taken into account. (This state of affairs is reminiscent of atomic spectroscopy: according to the classical laws of electrodynamics, an atom should have a continuous spectrum, whereas quantum mechanics allows only a discrete line emission.)
 
{\it Black-hole spectroscopy.---} In light of the preceding discussion, it is very suggestive to treat black holes as quantum objects, with a quantized surface area. The analogy with fundamental objects such as atoms \cite{Bekenmar} raises the possibility of associating black holes with other quantum phenomena. For instance, by analogy with atomic transitions, it is natural to ask whether there are selection rules which dictate the allowed black-hole quantum transitions. Perhaps such selection rules can be inferred from the black-hole QNM spectrum, utilizing Bohr's correspondence principle. 

In order to advocate this idea, we present a possible analogy between atomic spectroscopy and an intriguing feature of the black-hole QNM spectrum. Atomic transitions are constrained by the selection rule $\Delta l = \pm 1$ \cite{Ehren}, i.e. the angular momentum $l$ of the atom must change when emitting a radiation quantum. Somewhat similarly, the asymptotic ringing frequency of the (rotating) Kerr black hole {\it vanishes} when the azimuthal quantum number $m$ of the emitted field is zero \cite{Berti04}. Such a quasinormal mode corresponds to a virtual quantum-- it bears no energy. This, in turn, indicates that the corresponding quantum transition is forbidden. Hence, a Kerr black hole must change its angular momentum when emitting a field quantum, in close resemblance of the atom. 

Next, we point out that for a Schwarzschild black hole, the asymptotic real value of the QNM resonances is {\it zero} for fermionic fields \cite{Motl}, as opposed to the aforementioned $\ln3/4\pi$ value found for gravitational and scalar fields. Taking cognizance of the correspondence principle, 
this suggests that a quantized Schwarzschild black hole cannot emit a quantum with 
{\it half-integer} angular momentum-- such transitions seem to be forbidden. 

We conjecture that this selection-rule is more general, a genuine feature of black holes. In order to examine this hypothesis, we calculate the asymptotic ringing frequencies of a fermionic field in the RN spacetime.

The dynamics of a two-component Weyl neutrino field 
in the RN spacetime is governed by the Teukolsky wave equation \cite{Teukolsky,Note3,NoteMass}. 
The black hole QNMs correspond to solutions of the wave equation with the physical boundary
conditions of purely outgoing waves at spatial infinity and purely ingoing waves crossing the event horizon \cite{Detwe}. Such boundary conditions single out a discrete set of resonances $\{\omega_n\}$. 
The solution to the radial Teukolsky equation may be expressed as \cite{Leaver} 
(assuming an azimuthal dependence of the form $e^{im\phi}$)

\begin{eqnarray}\label{Eq4}
R_{lm}& = &e^{i\omega r} (r-r_-)^{-1-s+i\omega+i\sigma_+} (r-r_+)^{-s-i\sigma_+}\nonumber \\
&& \times \Sigma_{n=0}^{\infty} d_n \Big({{r-r_+} \over {r-r_-}}\Big)^n\  ,
\end{eqnarray}
where $r_{\pm} =M \pm (M^2-Q^2)^{1/2}$ are the black hole (event and inner) horizons, 
$\sigma_{+} \equiv \omega r_{+}/(r_{+}-r_{-})$, 
and the field spin-weight parameter $s$ takes the values $s= \pm {1 \over 2}$ for the 
neutrino field.

The sequence of expansion coefficients $\{d_n:n=1,2,3,\ldots\}$ is determined by a 
recurrence relation of the form \cite{Leaver}

\begin{equation}\label{Eq5}
\alpha_n d_{n+1}+\beta_n d_n +\gamma_n d_{n-1}=0\  ,
\end{equation}
with initial conditions $d_0=1$ and $\alpha_0 d_1+\beta_0 d_0=0$.
The quasinormal frequencies are determined by the requirement that the series 
in Eq. (\ref{Eq5}) is convergent, that is $\Sigma d_n$ exists and is finite \cite{Leaver}.

One finds \cite{HodKesh} that the physical content of the recursion coefficients becomes clear when they are expressed in terms of the Bekenstein-Hawking temperature $T_{BH}=(r_{+}-r_{-})/A$, where $A=4\pi r_+^2$ is the black-hole surface area. 
The recursion coefficients then obtain a surprisingly simple form, 

\begin{equation}\label{Eq6}
\alpha_n=(n+1)(n+1-s-2i\beta_{+}\omega)\  ,
\end{equation}

\begin{eqnarray}\label{Eq7}
\beta_n& = &-2(n+{1 \over 2}-2i\beta_{+}\omega)
(n+{1 \over 2}-2i\omega r_{+})\nonumber \\
&&-s-{1 \over 2} -A_{lm}\  ,
\end{eqnarray}
and
\begin{equation}\label{Eq8}
\gamma_n=(n-2i\omega)(n+s-2i\beta_{+}\omega)\  ,
\end{equation} 
where $\beta_{+} \equiv (4\pi T_{BH})^{-1}$ is the black-hole inverse temperature. 
The angular separation constants are given by $A_{lm}=l(l+1)-s(s+1)$, where $l \ge \max(|m|,|s|)$ is the angular momentum of the field \cite{Leaver}. We calculate numerically the quasinormal spectrum of neutrino fields by solving Eqs.~(\ref{Eq5})-(\ref{Eq8}) using the method of continued fractions \cite{Leaver,Nollert2}. 

We find that the neutrino quasinormal frequencies of the RN black hole exhibit a damped periodic behavior, where asymptotically $\omega_R \to 0$ as $|\omega_I| \to \infty$. 
Figure 1 demonstrates this damped periodic dependence of $\omega_R$ on $\omega_I$, in the complex frequency plane. In figure 2 we depict the envelope of these damped oscillations, which suggests that $\omega_R \propto \omega_I^{-1/2}$ in the asymptotic large damping limit. 
Our results thus confirm that the asymptotically vanishing resonances are a general feature of neutrino fields in spherically symmetric black-hole spacetimes. According to the correspondence principle, they describe virtual, zero energy quanta, suggesting that the corresponding quantum transitions are forbidden. 

\begin{figure}[tbh]
\centerline{\epsfxsize=9cm \epsfbox{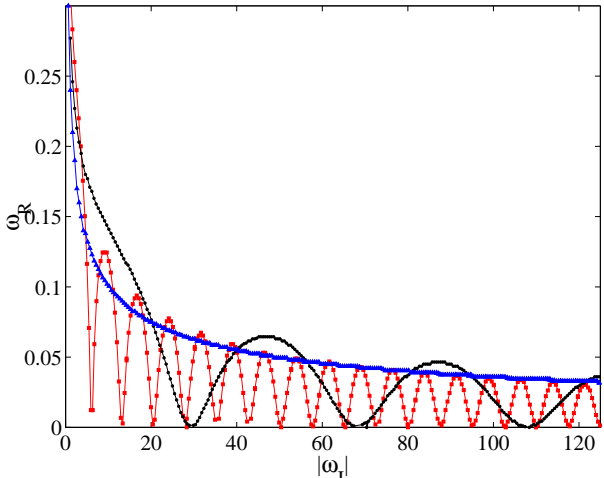}} 
\caption{Neutrino QNM frequencies of the RN black hole. Results are displayed for $l={1 \over 2}$, with $Q=0$ (triangles), $Q=0.3$ (circles) and $0.4$ (squares). Solid lines are a guide to the eye. The real part $\omega_R$ features a damped periodic dependence on the imaginary part $\omega_I$, with a period that becomes smaller with increasing black-hole charge. The spacing of the frequencies oscillates around $\Delta \omega_I=i2\pi T_{BH}$.}
\label{Fig1}
\end{figure}

\begin{figure}[tbh]
\centerline{\epsfxsize=9cm \epsfbox{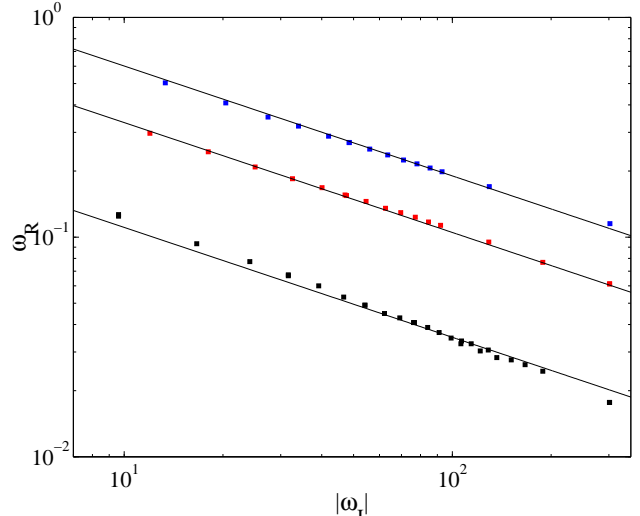}} 
\caption{Neutrino QNM frequencies of the RN black hole for $l={1 \over 2},{3 \over 2}$ and ${5 \over 2}$ (from bottom to top), with $Q=0.4$. For clarity, we display only the envelope of the $\omega_R(\omega_I)$ oscillatory dependence. The real part of the neutrino resonances asymptotically approaches zero, approximately as $\omega_R \propto \omega_I^{-1/2}$ (solid curves). }
\label{Fig2}
\end{figure}

{\it Summary.---} In the spirit of Bohr's correspondence principle, we suggest and demonstrate that asymptotically vanishing QNM frequencies may be interpreted as zero energy, forbidden quantum transitions. Motivated by this idea, we calculate the ringing frequencies of neutrino fields in the RN spacetime. It is shown that the spectrum of these black-hole resonances is characterized by a {\it vanishing} asymptotic real value. Our results raise the possibility that quantized spherically symmetric black holes cannot emit such fermionic quanta, possessing half-integer values of the angular momentum. 
Quantum black-hole selection rules, such as those demonstrated in this study, may provide an important clue in unveiling the underlying principles of the elusive theory of quantum gravity.

\bigskip
\noindent
{\bf ACKNOWLEDGMENTS}
\bigskip

The research of SH was supported by G.I.F. Foundation.


\begin{thebibliography}{99}
\bibitem{NoteMass} Note that in the large $|\omega|$ limit, the mass term is negligible with respect to the $\omega^2$ term in the wave equation. Thus, the mass of the field does not affect the asymptotic spectrum.
\end{thebibliography}
\end{document}